\pdfoutput=1
\pdfoutput=1
\pdfoutput=1
\pdfoutput=1
\pdfoutput=1
\RequirePackage{fix-cm}
\documentclass{svjour3}                     % onecolumn (standard format)
\smartqed  % flush right qed marks, e.g. at end of proof
\usepackage{graphicx,amsmath,amssymb,array}
\usepackage[utf8]{inputenc}
\begin{document}

\title{Adaptive Quantum Image Encryption Method Based on Wavelet Transform%\thanks{Grants or other notes
%about the article that should go on the front page should be
%placed here. General acknowledgments should be placed at the end of the article.}
}
%\subtitle{Do you have a subtitle?\\ If so, write it here}

%\titlerunning{Short form of title}        % if too long for running head

\author{Jian Wang \and Ya-Cong Geng \and Ji-Qiang Liu%etc.
}

%\authorrunning{Short form of author list} % if too long for running head

\institute{J. Wang \and Y.C. Geng\and J.Q. Liu \at
              Beijing Key Laboratory of Security and Privacy in Intelligent Transportation, Beijing Jiaotong University, Beijing 100044, China\\
              Science and Technology on Information Assurance Laboratory\\
              \email{wangjian@bjtu.edu.cn}           %  \\
%             \emph{Present address:} of F. Author  %  if needed
%           \and
%           S. Author \at
%              second address
}

\date{Received: date / Accepted: date}
% The correct dates will be entered by the editor

\maketitle

\begin{abstract}
An adaptive quantum image encryption method based on wavelet transform is designed. Since the characteristic of most information is centralized in the low frequency part after performing the wavelet transform, it reserves the image low frequency information only, so as to reduce the encryption workload. Then it encrypts the low frequency information by the random key stream generated by logistic map, the encryption process is realized by implementing XOR operation. In the decryption process, it carries out zero filling operations for high frequency coefficient to recover the decryption images which are equal to the plain images. At the same time, the relevant quantum logical circuit is designed. Statistical simulation and theoretical analysis demonstrate that the proposed quantum image encryption algorithm has higher security and lower computational complexity. So it is adaptive to the scenes that need to encrypt a large number of images during network transmission.
% \PACS{PACS code1 \and PACS code2 \and more}
% \subclass{MSC code1 \and MSC code2 \and more}
\end{abstract}

\section{Introduction}
With the development of internet technology, as one of the main information carriers on the network, image may includes a lot of individual private information and important data, therefore,  how to protect the safety of image effectively has become the more and more important problem. And at the same time, a lot of classical image encryption schemes are put forward. However, not only an image has the mass data, but also the total quantity of the images on the network is also very large in this era of big data, therefore, how to encrypt the images safely and high efficiently is very important. 

In recent years, quantum computer has been known for its ultra-high computing capacity. It achieves the computing with high degree of parallelism by the characteristics of superposition and entanglement. This is undoubted to bring a new research idea to image encryption with large data, hence more and more researches focus on the quantum image encryption. The introduction of the existing main search results are shown as below.

In 2013, Zhou R.G. et al. \cite{1} put forward the quantum image encryption and decryption algorithms based on quantum image geometric transformations, and at the same time designed the relevant quantum circuits. This algorithm applies quantum image geometric transformation, which provides the new idea for the research of quantum image cryptography.
 
In the same year, Ahmed et al. \cite{2} presented a novel color image chaotic encryption algorithm based on quantum chaotic system, which applied the chaotic system to the image encryption technology at the first time, provides the new direction for the research of image encryption.

In 2015, Zhou N.R. et al. \cite{3} designed the quantum implementation of the generalized Arnold transform. Put forward a kind of quantum image encryption algorithm based on the generalized Arnold transform and double random phase encoding operationgs. Carried out the scrambling for the pixels by the generalized Arnold transform, and carried out the encoding for the gray-scale information of the image by double random phase operations. This algorithm has the better effectiveness, and its computing complexity is $O\left( {{n}^{2}} \right)$, much lower than the corresponding classical algorithm.

In 2016, Tan R.C. et al. \cite{4} designed a quantum color image encryption algorithm based on a hyper-chaotic system and quantum Fourier transform, in which the hyper-chaotic sequences are scrambled with three components of the original color image. Then the quantum Fourier transform is performed to fulfill the encryption. It verified that the computing complexity of the algorithm is $O\left( {{n}^{2}} \right)$. 

In 2018, Ran Q. et al. \cite{5} put forward a kind of quantum color image encryption scheme based on coupled hyper-chaotic Lorenz system with three impulse injections. In order to strength the complexity of orbit, three pulse signals are injected to the coupled hyper-chaotic Lorenz system. And six sequences generated by this system are to encrypt the quantum color image by XOR operation and right cyclic shift operation. The proposed encryption scheme has the good feasibility and effectiveness on the encryption quantum color image.

In the same year, Liu X.B. et al.\cite{6} put forward a kind of quantum gray-scale image encryption algorithm based on QAT (quantum Arnold transform) and quantum random rotation, which encrypted the quantum gray-scale images by the combination of quantum permutation and quantum bit random rotation, and discussed the computing complexity of algorithm was $O\left( {{n}^{2}} \right)$.

In 2019, Jiang N. et al.\cite{7} designed a quantum image encryption based on Henon mapping, it breaked away from the restriction of classical computers and did all the works in quantum computers, including the generation of the chaos sequences, the encryption and decryption processes. And experiments shows that the algorithm is safe and reliable, but it still has some shortcomings in efficiency.

This paper divides the research direction of quantum image encryption into two types. One is based on quantum transformation and random rotation. This kind of research achieves the purpose of encryption through various transformations or quantum rotations on image position information and color information, such as references \cite{1,3,6} and \cite{8,9,10,11,12}, this research type is the main direction of the research at present. The other is based on the chaotic systems. This kind of research manipulates the quantum basic gates through a series of chaotic sequences generated by the chaotic systems, so as to realize the encryption of quantum images, such as references \cite{2,4,5} and \cite{13}, the research of this type is still in the start-up stage.

Through the above analysis of the complexity of the existing quantum image encryption algorithms, it can be seen that compared with the classical image encryption algorithms, the encryption efficiency of the quantum one has been greatly improved, which is due to the unique characteristics of quantum computing model, such as superposition and entanglement. These characteristics greatly improve the efficiency of complex image processing algorithms. However, on the basis of quantum model, the encryption efficiency of the existing quantum image encryption algorithms is not very ideal, and there are two reasons.

(1) The researches based on quantum transformation and random rotation need to performing complex transformations and rotations for many times to enhance the security of images, which leads to the increasement of complexity. Therefore, the computational complexity of the most researches is still $O\left( {{n}^{2}} \right)$, some even reach the exponential level.

(2) The researches based on chaotic systems do not need to rely on the transformations, which is similar to the "one-time pad" system. However, they don’t make full use of the huge advantages brought by quantum characteristics. As a result, the encryption efficiency of this type has not been greatly improved compared with the former. Although the complexity of a few studies has reached linear level, the coefficients are still very large.

In order to explore more high efficient quantum image encryption methods, this paper introduces the wavelet transform from the classical image encryption field, it tries to reduce the encryption computing quantity on the basis of ensuring the security of the algorithm. The general structure of this paper shows: 1. Brief conclusion of the current quantum image encryption algorithm; 2. Introduction of the relevant knowledge; 3. Design of the quantum image encryption and decryption algorithms based on wavelet transform and the implementation of its circuit; 4. Analysis and evaluation of the experiment; 5. Conclusion of the paper.

\section{Fundamental Knowledge}

\subsection{The Generalized Quantum Image Representation}
In this algorithm, the image to be encrypted is represented by the generalized quantum image representation (GQIR) model \cite{14}. The GQIR stores the position information and color information of a $H\times W$ image by two entangled quantum bit sequences.

Assume that the gray-scale value of an image is within $0\sim {{2}^{q}}-1$, and then the binary sequence $C_{YX}^{0}C_{YX}^{1}...C_{YX}^{q-2}C_{YX}^{q-1}$ is used to encode the color information $\left| {{C}_{YX}} \right\rangle $. The position information$\left| YX \right\rangle $ is encoded by $\left| {{y}_{0}}{{y}_{1}}...{{y}_{h-1}} \right\rangle \left| {{x}_{0}}{{x}_{1}}...{{x}_{w-1}} \right\rangle $ , where
\begin{equation}
h=\left\{ \begin{array}{*{35}{l}}
\left\lceil {{\log }_{2}}H \right\rceil \text{ },\text{    }H>1  \\
1\text{ },\text{    }H=1  \\
\end{array} \right.
\end{equation}
\begin{equation}
w=\left\{ \begin{array}{*{35}{l}}
\left\lceil {{\log }_{2}}W \right\rceil \text{ },\text{    }W>1  \\
1\text{ },\text{    }W=1  \\
\end{array} \right.
\end{equation}
$h$ and $w$ are integers.

Then the $H\times W$ quantum image based on GQIR can be written as below.
\begin{equation}
\begin{aligned}
	\left| I \right\rangle &=\frac{1}{{\sqrt{2}^{h+w}}}\sum\limits_{Y=0}^{H-1}{\sum\limits_{X=0}^{W-1}{\underset{i=0}{\overset{q-1}{\mathop{\otimes }}}\,\left| C_{YX}^{i} \right\rangle }}\left| YX \right\rangle  \\ 
&=\frac{1}{{\sqrt{2}^{h+w}}}\sum\limits_{Y=0}^{H-1}{\sum\limits_{X=0}^{W-1}{\left| f\left( Y,X \right) \right\rangle }}\left| YX \right\rangle   
\end{aligned}
\end{equation}

\subsection{Wavelet Transform}
In the classical field, the definition of the constant wavelet transform \cite{15} is:
\begin{equation}
{{W}_{\psi }}f(a,b)=\int_{-\infty }^{+\infty }{f(t)\psi _{a,b}^{*}(t)dt=\left\langle f(t),{{\psi }_{a,b}}(t) \right\rangle }
\end{equation}
where, window function is ${{\psi }_{a,b}}(t)={{\left| a \right|}^{-\frac{1}{2}}}\psi (\frac{t-b}{a})$, $\psi (t)$is wavelet basis function, scale parameter $a\in R$ and $a\ne 0$, $b$ is position parameter; $\psi _{a,b}^{*}(t)$ is the conjugate of ${{\psi }_{a,b}}(t)$. The reverse transform $f(t)$ of constant wavelet transform is:
\begin{equation}
f(t)=\frac{1}{{{C}_{\psi }}}\int_{-\infty }^{+\infty }{\int_{-\infty }^{+\infty }{{{a}^{-2}}{{W}_{\psi }}f(a,b){{\psi }_{a,b}}(t)dadb}}
\end{equation}
\begin{equation}
{{C}_{\psi }}=\int_{\text{-}\infty }^{+\infty }{\frac{{{\left| {{\psi }^{V}}(w) \right|}^{\text{2}}}}{w}dw<\infty ,{{\psi }^{V}}(w)=\int_{-\infty }^{+\infty }{\psi (t){{e}^{-jwt}}dt}}
\end{equation}
$w$ is the frequency, $t$ is the time.

Meanwhile, in the quantum field, Hoyer \cite{16} gave out the definitions of several wavelet transforms. Among them, Daubechies-D(4) wavelet transform matrix was shown as formula (7), ${{P}_{{{\text{2}}^{n}}}}$ is the permutation matrix.
\begin{equation}
D_{{{2}^{n}}}^{(4)}=\left( {{I}_{{{2}^{n-1}}}}\otimes {{C}_{1}} \right){{P}_{{{2}^{n}}}}\left( {{I}_{{{2}^{n-1}}}}\otimes {{C}_{0}} \right)
\end{equation}
In the formula, ${{C}_{\text{1}}}=\frac{\text{1}}{\text{2}}\left[ \begin{matrix}
\frac{{{c}_{0}}}{{{c}_{3}}} & 1  \\
1 & \frac{{{c}_{1}}}{{{c}_{2}}}  \\
\end{matrix} \right]\text{ },\text{   }{{C}_{\text{0}}}=\left[ \begin{matrix}
{{c}_{3}} & -{{c}_{2}}  \\
{{c}_{2}} & {{c}_{3}}  \\
\end{matrix} \right]$, the coefficients are ${{c}_{0}}=\frac{1+\sqrt{3}}{2\sqrt{2}},\text{  }{{c}_{1}}=\frac{3+\sqrt{3}}{2\sqrt{2}},\text{  }{{c}_{2}}=\frac{3-\sqrt{3}}{2\sqrt{2}},\text{  }{{c}_{3}}=\frac{1-\sqrt{3}}{2\sqrt{2}}$.

The wavelet transform is the effective tool of digital image processing, its first-class decomposition effect is shown as Fig. 1, LL, HL, LH, HH are respectively low frequency component, horizontal edge details, vertical edge details and diagonal direction details.

\begin{figure}[h]
	\centering
	\includegraphics[width=0.8\textwidth]{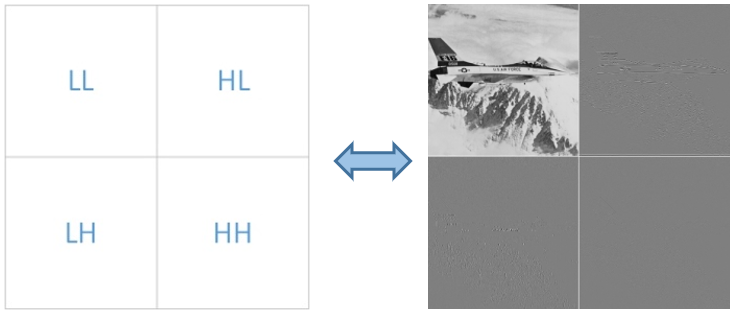}
	\caption{Wavelet first-class decomposition}
	\label{fig:1}
\end{figure}

\subsection{Analysis on the Impact of Low Frequency Coefficient}
From Fig. 1, we can see that the low frequency coefficient of an image includes most information of the image, it is equivalent to a thumbnail of the original image. In order to further explore the impact of low frequency coefficient on the entire image, we conducted the following experiments, the experiment steps are shown as below. 

Step 1: Perform Daubechies-D(4) wavelet transform for a $\text{512}\times \text{512}$ image, decompose the image to four parts: low frequency component, horizontal edge details, vertical edge details and diagonal direction details.

Step 2: Make 0,1 substitution processing for the low frequency component, and remain the high frequency component (horizontal edge details, vertical edge details and diagonal direction details) unchanged.

Step 3: Perform the inverse wavelet transform for the processed image, and the experiment result diagrams are shown as Fig. 2.

\begin{figure}[!h]
	\centering
	\begin{minipage}{0.32\textwidth}
		\centering
		\includegraphics[width=\textwidth]{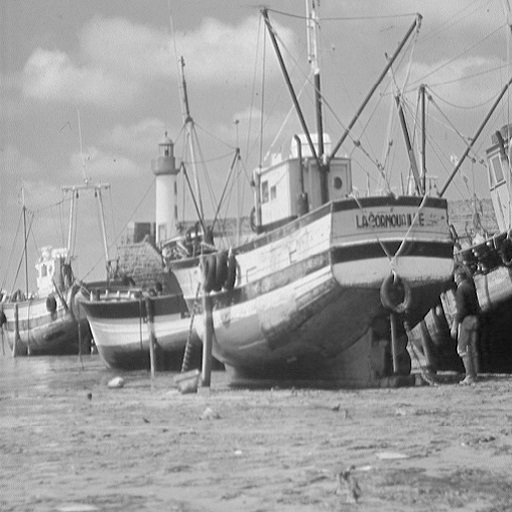}\\
		(1)
	\end{minipage}
	\begin{minipage}{0.32\textwidth}
		\centering
		\includegraphics[width=\textwidth]{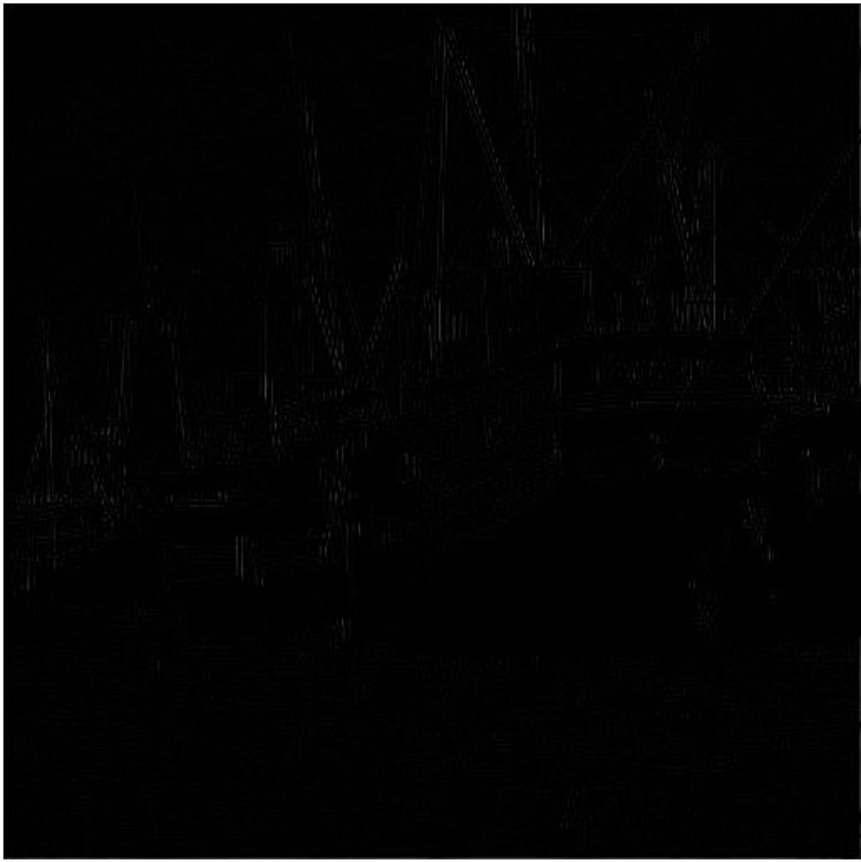}\\
		(2)
	\end{minipage}
	\begin{minipage}{0.32\textwidth}
		\centering
		\includegraphics[width=\textwidth]{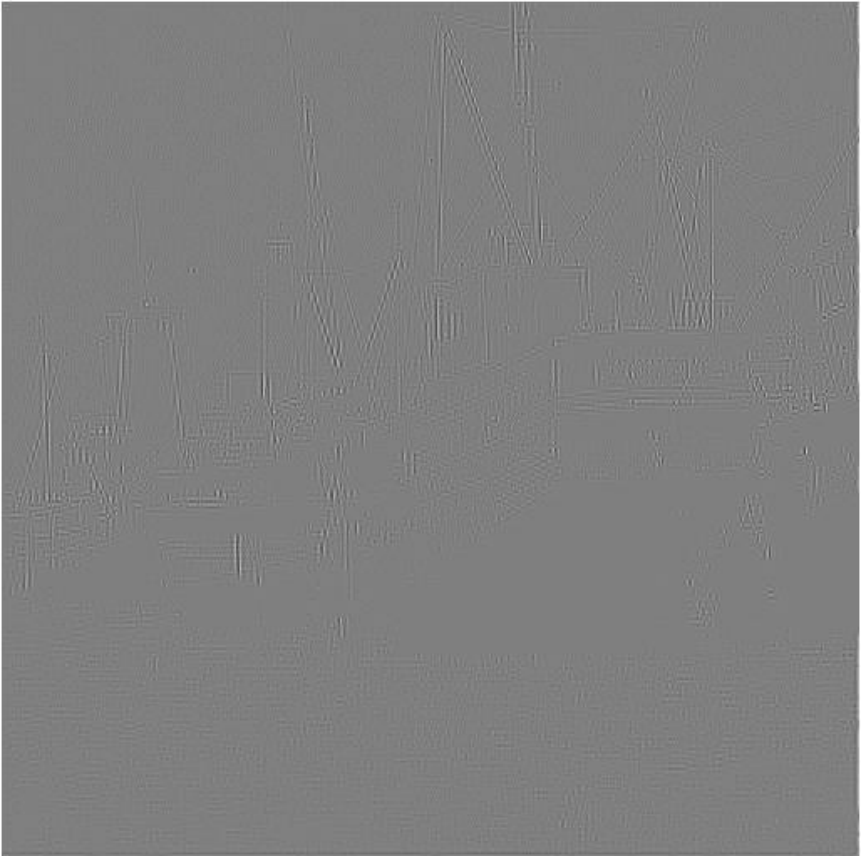}\\
		(3)
	\end{minipage}
	\caption{(1) Plain image of “boat”; (2) The image of “boat” after making 0 substitution processing for the low frequency coefficient;  (3) The image of “boat” after making 1 substitution processing for the low frequency coefficient
	}
\end{figure}

From Fig. 2, we can see that we basically can not find out the original image information from the recovered image after carried out 0, 1 substitution processing for the low frequency component, it only exists a small amount of details. So it follows that the most important part of an image is the low frequency part, which is crucial for the whole image, while the high frequency part is the detailed information of the image, which has little effect on the whole image. In addition, human’s vision system is limited by physiological characteristics, they pay uneven attentions to the image field, and don't have the obvious feeling for the subtle difference in details. Based on the above two points, wavelet transform is widely used to image compression, image watermarking and image encryption in the classical field.

\section{Quantum Image Encryption and Decryption Algorithm based on Wavelet Transform}

In 1998, Amir Fijany et al. \cite{17} researched several kinds of quantum circuit models of wavelet transform, which provided the technical supports for the wavelet transform application in quantum field, meanwhile, we call this kind of wavelet transform is quantum wavelet transform (QWT). Based on this, using the characteristic that most information of the image after wavelet transform are all reserved in low frequency part, this paper puts forward an adaptive quantum image encryption algorithm. This algorithm just reserves the low frequency information of image, greatly reduced the encryption workload, as well as the expenses on image transmission. It makes zero filling operation for high frequency coefficient to recover the decryption images that are equal to the original images. The overall framework of the encryption and decryption algorithms are shown as Fig. 3. 
\begin{figure}[h]
	\centering
	\includegraphics[width=1\textwidth]{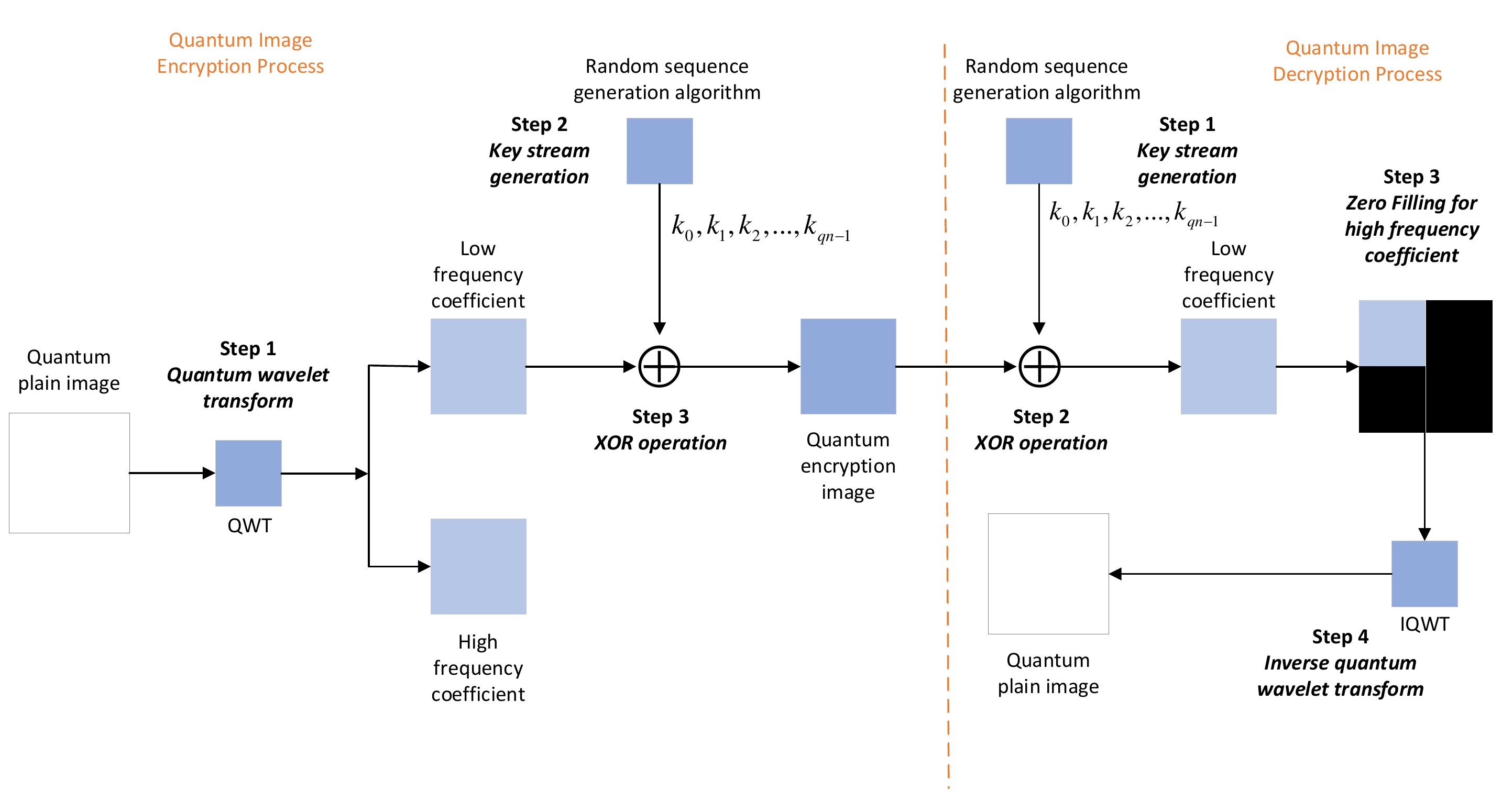}
	\caption{Overall framework of the encryption and decryption algorithms}
	%\label{fig:3}
\end{figure}
\subsection{Quantum Image Encryption Process Based on Wavelet Transform}

As shown in Fig. 3, the quantum image encryption process based on wavelet transform can be divided into three steps: 1. Perform wavelet transform for the original image and get its low frequency coefficient; 2. Generate key stream by random sequence generation algorithm; 3. Carry out XOR operation with key stream and low frequency part. The specific operating steps are show as below.

Step 1: Quantum wavelet transform

Firstly, carry out QWT \cite{17} for the original quantum image $\left| I \right\rangle $, and get its quantum wavelet coefficient.
\begin{equation}
\begin{aligned}
 QWT\left| I \right\rangle &=\frac{1}{{\sqrt{2}^{h+w}}}\sum\limits_{Y=0}^{H-1}{\sum\limits_{X=0}^{W-1}{QWT\left| f\left( Y,X \right) \right\rangle }}\left| YX \right\rangle  \\ 
& =\frac{1}{{\sqrt{2}^{h+w}}}\sum\limits_{Y=0}^{{{2}^{h\text{-1}}}-1}{\sum\limits_{X=0}^{{{2}^{w\text{-1}}}-1}{\left| tf\left( Y,X \right) \right\rangle }}\left| YX \right\rangle \\ &+\frac{1}{{\sqrt{2}^{h+w}}}\sum\limits_{Y=0}^{{{2}^{h-1}}-1}{\sum\limits_{X={{2}^{w-1}}}^{{{2}^{w}}-1}{\left| tf\left( Y,X \right) \right\rangle }}\left| YX \right\rangle  \\ 
& +\frac{1}{{\sqrt{2}^{h+w}}}\sum\limits_{Y={{2}^{h-1}}}^{{{2}^{h}}-1}{\sum\limits_{X=0}^{{{2}^{w-1}}-1}{\left| tf\left( Y,X \right) \right\rangle }}\left| YX \right\rangle  \\ &+\frac{1}{{\sqrt{2}^{h+w}}}\sum\limits_{Y={{2}^{h-1}}}^{{{2}^{h}}-1}{\sum\limits_{X={{2}^{w-1}}}^{{{2}^{w}}-1}{\left| tf\left( Y,X \right) \right\rangle }}\left| YX \right\rangle  \\ 
& =\left| {{I}_{\text{1}}} \right\rangle +\left| {{I}_{\text{2}}} \right\rangle   
\end{aligned}
\end{equation}
In the formula (8), $\left| tf\left( Y,X \right) \right\rangle $ represents the gray value of the image after performing QWT. And then, the low frequency information of original image shows $\left| {{I}_{\text{1}}} \right\rangle $, the remaining high frequency information shows $\left| {{I}_{\text{2}}} \right\rangle $, where $\left| {{I}_{\text{1}}} \right\rangle $is:
\begin{equation}
\begin{aligned}
\left| {{I}_{\text{1}}} \right\rangle &=\frac{1}{{\sqrt{2}^{h+w}}}\sum\limits_{Y=0}^{{{2}^{h\text{-1}}}-1}{\sum\limits_{X=0}^{{{2}^{w\text{-1}}}-1}{\left| tf\left( Y,X \right) \right\rangle }}\left| YX \right\rangle  \\ 
&=\frac{1}{{\sqrt{2}^{h+w}}}\sum\limits_{Y=0}^{{{2}^{h\text{-1}}}-1}{\sum\limits_{X=0}^{{{2}^{w\text{-1}}}-1}{\left| g\left( Y,X \right) \right\rangle }}\left| YX \right\rangle   
\end{aligned}
\end{equation}
 $\left| g\left( Y,X \right) \right\rangle $ is the same as $\left| tf\left( Y,X \right) \right\rangle $.
 
 Therefore, low frequency part can be stored by $h+w-2+q$ quantum bits, where $h+w-2$ qubits store position information, $q$ qubits store color information.
\begin{equation}
\left| {{I}_{\text{1}}} \right\rangle =\frac{1}{{\sqrt{2}^{{h}'+{w}'}}}\sum\limits_{Y=0}^{{{2}^{{{h}'}}}-1}{\sum\limits_{X=0}^{{{2}^{{{w}'}}}-1}{\left| g\left( Y,X \right) \right\rangle }}\left| YX \right\rangle 
\end{equation}
In the formula(10), ${h}'=h-1,{w}'=w-1$.

Step 2: Key stream generation

This paper doesn't limit the encryption algorithm used to generate the key stream. Any one that generates a random sequence can be used to generate the key stream. Assume the key stream is $K=\left\{ {{k}_{0}},{{k}_{1}},{{k}_{2}},...,{{k}_{qn-1}} \right\}$, where ${{k}_{i}}\in \left\{ 0,1 \right\}$, $i\in \left[ 0,qn-1 \right]$, 
$q$ is the number of qubits needed to store image color information, $n={HW}/{4}\;$ is the length of key stream, which is 1/4 of the size of plain image.

Step 3: XOR operation (Low frequency information encryption)

Due to the low frequency coefficient obtained from the step 1 includes most information of image, we do XOR operation between low frequency coefficient and key stream. Set XOR operation as $B$[5] , the encrypted image as $\left| Q \right\rangle $, and then:
\begin{equation}
\begin{aligned}
\left| Q \right\rangle 
&=B\left| {{I}_{1}} \right\rangle  \\ 
&=\prod\limits_{Y=0}^{{{2}^{{{h}'}}}-1}{\prod\limits_{X=0}^{{{2}^{{{w}'}}}-1}{{{B}_{YX}}\left| {{I}_{1}} \right\rangle }} \\ 
&=\frac{1}{{\sqrt{2}^{{h}'+{w}'}}}\sum\limits_{Y=0}^{{{2}^{{{h}'}}}-1}{\sum\limits_{X=0}^{{{2}^{{{w}'}}}-1}{\underset{i=0}{\overset{q-1}{\mathop{\otimes }}}\,\left| C_{YX}^{i}\oplus k_{YX}^{i} \right\rangle }}\left| YX \right\rangle  \\ 
&=\frac{1}{{\sqrt{2}^{{h}'+{w}'}}}\sum\limits_{Y=0}^{{{2}^{{{h}'}}}-1}{\sum\limits_{X=0}^{{{2}^{{{w}'}}}-1}{\left| m(Y,X) \right\rangle }}\left| YX \right\rangle   
\end{aligned}
\end{equation}
where, $\left| m\left( Y,X \right) \right\rangle $ represents the color information of low frequency part after doing XOR operation. ${{B}_{YX}}$ is the suboperation of $B$, it shows as (12).
\begin{equation}
\begin{aligned}
 {{B}_{YX}}\left| {{I}_{1}} \right\rangle 
 &={{B}_{YX}}\left( \frac{1}{{\sqrt{2}^{{h}'+{w}'}}}\sum\limits_{y=0}^{{{2}^{{{h}'}}}-1}{\sum\limits_{x=0}^{{{2}^{{{w}'}}}-1}{\underset{i=0}{\overset{q-1}{\mathop{\otimes }}}\,\left| C_{yx}^{i} \right\rangle }}\left| yx \right\rangle  \right) \\ 
& =\frac{1}{{\sqrt{2}^{{h}'+{w}'}}}{{B}_{YX}}\left( \sum\limits_{y=0}^{{{2}^{{{h}'}}}-1}{\sum\limits_{\begin{smallmatrix} 
		x=0 \\ 
		yx\ne YX 
		\end{smallmatrix}}^{{{2}^{{{w}'}}}-1}{\underset{i=0}{\overset{q-1}{\mathop{\otimes }}}\,\left| C_{yx}^{i} \right\rangle }}\left| yx \right\rangle +\underset{i=0}{\overset{q-1}{\mathop{\otimes }}}\,\left| C_{YX}^{i} \right\rangle \left| YX \right\rangle  \right) \\ 
& =\frac{1}{{\sqrt{2}^{{h}'+{w}'}}}\left( \sum\limits_{y=0}^{{{2}^{{{h}'}}}-1}{\sum\limits_{\begin{smallmatrix} 
		x=0 \\ 
		yx\ne YX 
		\end{smallmatrix}}^{{{2}^{{{w}'}}}-1}{\underset{i=0}{\overset{q-1}{\mathop{\otimes }}}\,\left| C_{yx}^{i} \right\rangle }}\left| yx \right\rangle +\underset{i=0}{\overset{q-1}{\mathop{\otimes }}}\,\left| C_{YX}^{i}\oplus k_{YX}^{i} \right\rangle \left| YX \right\rangle  \right)  
\end{aligned}
\end{equation}

\subsection{Quantum Image Decryption Process Based on Wavelet Transform}
As shown in Fig. 3, quantum image decryption process based on wavelet transform can be divided into four steps: 1. Generate the key stream sequence used for decryption; 2. Carry out XOR operation with key stream and encryption image to recover the low frequency part of original image; 3. Fill the high frequency part with zero to get the middle image; 4. Perform the wavelet transform for the middle image and get the decrypted image. The specific operating steps are shown as below:

Step 1: Key stream generation

The decrypted key stream $K=\left\{ {{k}_{0}},{{k}_{1}},{{k}_{2}},...,{{k}_{qn-1}} \right\}$ is obtained by the random sequence same as the encryption process, where ${{k}_{i}}\in \left\{ 0,1 \right\}$, $i\in \left[ 0,qn-1 \right]$, $n=H\times W$ is the length of key stream.

Step 2: XOR operation (Low frequency information decryption)

Carry out $B$ operation for the encrypted information $\left| Q \right\rangle $, and get the decrypted low frequency information $\left| {{I}_{\text{1}}} \right\rangle $.
\begin{equation}
\begin{aligned}
\left| {{I}_{\text{1}}} \right\rangle 
&=B\left| Q \right\rangle  \\ 
& =\prod\limits_{Y=0}^{{{2}^{{{h}'}}}-1}{\prod\limits_{X=0}^{{{2}^{{{w}'}}}-1}{{{B}_{YX}}\left| Q \right\rangle }} \\ 
& =\frac{1}{{\sqrt{2}^{{h}'+{w}'}}}\sum\limits_{Y=0}^{{{2}^{{{h}'}}}-1}{\sum\limits_{X=0}^{{{2}^{{{w}'}}}-1}{\underset{i=0}{\overset{q-1}{\mathop{\otimes }}}\,\left| {C_{YX}^{i}}'\oplus k_{YX}^{i} \right\rangle }}\left| YX \right\rangle  \\ 
& =\frac{1}{{\sqrt{2}^{{h}'+{w}'}}}\sum\limits_{Y=0}^{{{2}^{{{h}'}}}-1}{\sum\limits_{X=0}^{{{2}^{{{w}'}}}-1}{\left| g(Y,X) \right\rangle }}\left| YX \right\rangle   
\end{aligned}
\end{equation}

Step 3: Zero filling for high frequency coefficient

Due to the encryption process only reserves the low frequency information of the original image, and then the image recovered from Step 2 is just 1/4 of original image, for getting the decrypted image is the same size as the original image, we make zero filling processing for the high frequency part, so as to get the middle image $\left| \psi  \right\rangle $.
\begin{equation}
\begin{aligned}
\left| \psi  \right\rangle &=\frac{1}{{\sqrt{2}^{h+w}}}\sum\limits_{Y=0}^{{{2}^{h\text{-1}}}-1}{\sum\limits_{X=0}^{{{2}^{w\text{-1}}}-1}{\left| {{g}_{1}}\left( Y,X \right) \right\rangle }}\left| YX \right\rangle  \\ 
&
+\frac{1}{{\sqrt{2}^{h+w}}}\sum\limits_{Y=0}^{{{2}^{h-1}}-1}{\sum\limits_{X={{2}^{w-1}}}^{{{2}^{w}}-1}{\left| {{g}_{2}}\left( Y,X \right) \right\rangle }}\left| YX \right\rangle  \\ 
& +\frac{1}{{\sqrt{2}^{h+w}}}\sum\limits_{Y={{2}^{h-1}}}^{{{2}^{h}}-1}{\sum\limits_{X=0}^{{{2}^{w-1}}-1}{\left| {{g}_{3}}\left( Y,X \right) \right\rangle }}\left| YX \right\rangle \\
&
+\frac{1}{{\sqrt{2}^{h+w}}}\sum\limits_{Y={{2}^{h-1}}}^{{{2}^{h}}-1}{\sum\limits_{X={{2}^{w-1}}}^{{{2}^{w}}-1}{\left| {{g}_{4}}\left( Y,X \right) \right\rangle }}\left| YX \right\rangle  \\ 
& =\frac{1}{{\sqrt{2}^{h+w}}}\sum\limits_{Y=0}^{{{2}^{h}}-1}{\sum\limits_{X=0}^{{{2}^{w}}-1}{\left| {g}'\left( Y,X \right) \right\rangle }}\left| YX \right\rangle   
\end{aligned}
\end{equation}
In the formula, $\left| {{g}_{1}}\left( Y,X \right) \right\rangle $ represents the recovered low frequency coefficient value; $\left| {{g}_{\text{2}}}\left( Y,X \right) \right\rangle $, $\left| {{g}_{\text{3}}}\left( Y,X \right) \right\rangle $, $\left| {{g}_{\text{4}}}\left( Y,X \right) \right\rangle $represent the high frequency coefficient value, they are ${{\left| \text{0} \right\rangle }^{\otimes q}}$;  $\left| {g}'\left( Y,X \right) \right\rangle $ represents the color information of the middle image $\left| \psi  \right\rangle $.

\begin{figure}[h]
	\centering
	\includegraphics[width=0.4\textwidth]{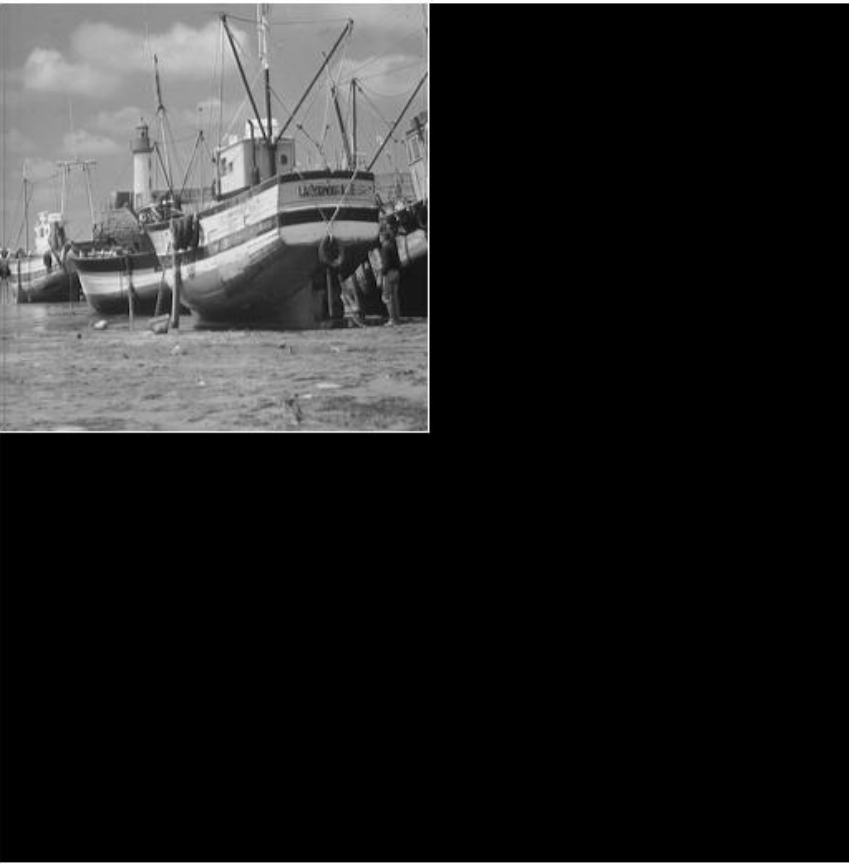}
	\caption{Example of middle image $\left| \psi  \right\rangle $: zero filling for high frequency part}
	\label{fig:4}
\end{figure}

Step 4: Inverse quantum wavelet transform 

Finally, we need to perform the inverse quantum wavelet transform (IQWT) for the middle image $\left| \psi  \right\rangle $ obtained from Step 3, and get the final decrypted image $\left| {{I}'} \right\rangle $.
\begin{equation}
\begin{aligned}
IQWT\left( \left| \psi  \right\rangle  \right)
&=\frac{1}{{\sqrt{2}^{h+w}}}\sum\limits_{Y=0}^{{{2}^{h}}-1}{\sum\limits_{X=0}^{{{2}^{w}}-1}{IQWT\left| {g}'\left( Y,X \right) \right\rangle }}\left| YX \right\rangle  \\ 
& =\frac{1}{{\sqrt{2}^{h+w}}}\sum\limits_{Y=0}^{{{2}^{h}}-1}{\sum\limits_{X=0}^{{{2}^{w}}-1}{\left| {t}'{g}'\left( Y,X \right) \right\rangle }}\left| YX \right\rangle  \\ 
& =\frac{1}{{\sqrt{2}^{h+w}}}\sum\limits_{Y=0}^{{{2}^{h}}-1}{\sum\limits_{X=0}^{{{2}^{w}}-1}{\left| {f}'\left( Y,X \right) \right\rangle }}\left| YX \right\rangle  \\ 
& =\left| {{I}'} \right\rangle   
\end{aligned}
\end{equation}

Where, $\left| {t}'{g}'\left( Y,X \right) \right\rangle $ represents the color information after performing IQWT, which is the same as $\left| {f}'\left( Y,X \right) \right\rangle $.
\subsection{Design of Quantum Circuit} 

It can be seen from Fig. 3 that the modules of the encryption and decryption processes mainly involve QWT, obtaining the low frequency coefficient, XOR operation, zero filling for high frequency coefficient and IQWT, where, the circuits of QWT and XOR operation have been given out in Ref. \cite{17,7}, the circuit of IQWT is similar to QWT, so we will not state these circuits here. This section will focus on the circuits of obtaining of low frequency coefficient and zero filling for high frequency coefficient, then give the overall circuits of the encryption and decryption processes.

\subsubsection{Circuit of obtaining the low frequency coefficient}

Fig. 5 shows the circuit of obtaining the low frequency part of a ${{\text{2}}^{h}}\times {{2}^{w}}$ image after the wavelet transform. Due to the low frequency is reserved in low position, the highest position on the horizontal, vertical coordinate are zero, namely every position shall be meet $\left| {{X}_{\text{0}}} \right\rangle =\left| \text{0} \right\rangle \text{ }\left| {{Y}_{\text{0}}} \right\rangle =\left| \text{0} \right\rangle $, such position has ${{\text{2}}^{h-1}}\times {{2}^{w-1}}$ in total. In the figure, we use ETOF gate \cite{18,19} (the extended Toffoli gate) to select the coefficient on every position that meets the condition. After the coefficient on this position is selected, it will be turned over, therefore, we turn over again with an auxiliary quantum bit $\left| A \right\rangle $, so as to keep the final obtained coefficient consistent with the coefficient of original image. Each block in the figure represents the coefficient on the selected position. 

\begin{figure}[h]
	\centering
	\includegraphics[width=1\textwidth]{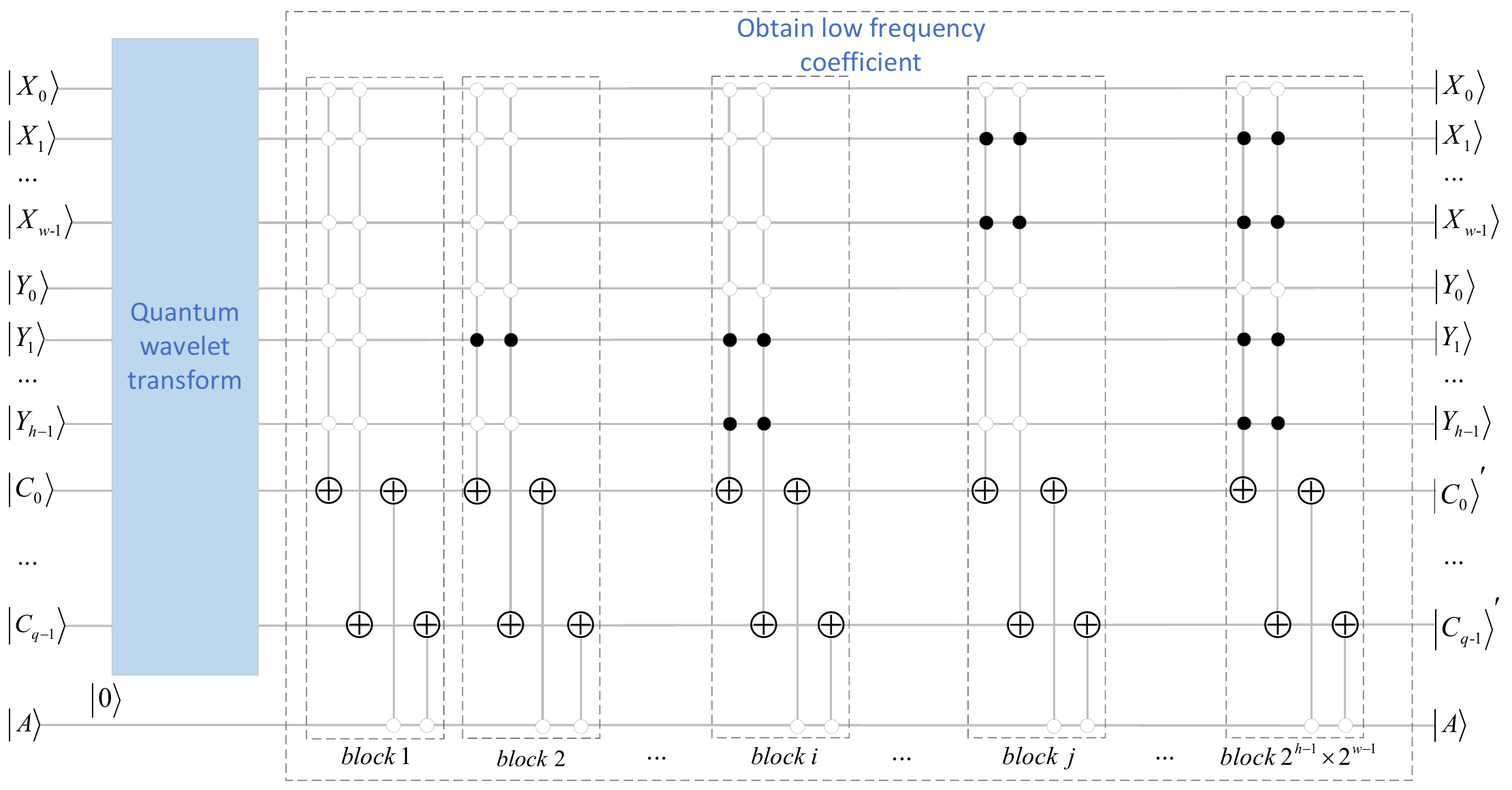}
	\caption{Circuit of obtaining the low frequency coefficient }
	\label{fig:5}
\end{figure}

Fig. 6 gives an example. For a image size of ${{\text{2}}^{\text{2}}}\times {{\text{2}}^{\text{2}}}$, its low frequency coefficients are existed on $\left| \text{0000} \right\rangle $, $\left| \text{0100} \right\rangle $, $\left| \text{0001} \right\rangle $and $\left| \text{0101} \right\rangle $, so, we get the low frequency information with four blocks.

\begin{figure}[h]
	\centering
	\includegraphics[width=0.8\textwidth]{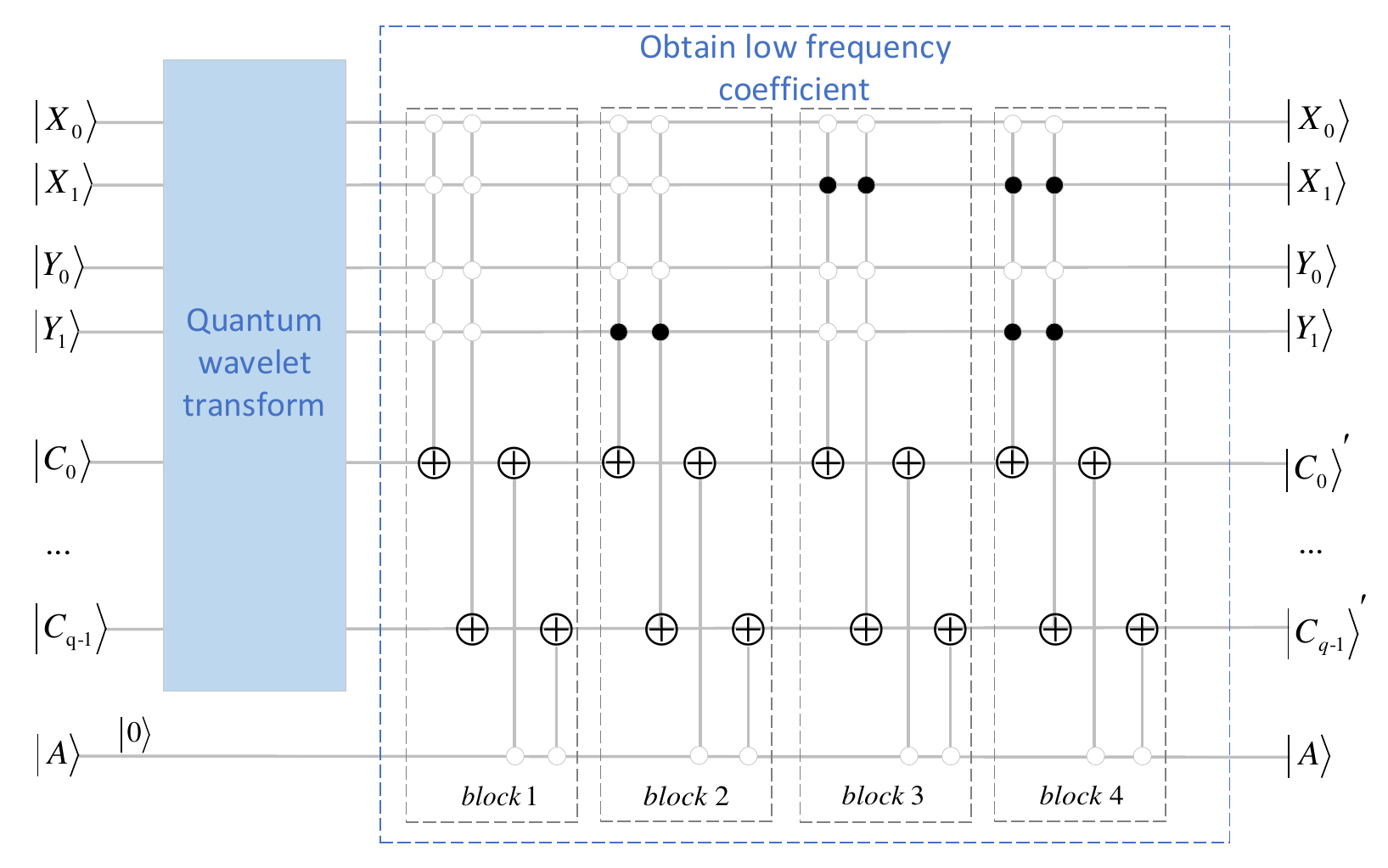}
	\caption{Example of obtaining the low frequency coefficient of ${{\text{2}}^{\text{2}}}\times {{\text{2}}^{\text{2}}}$ image }
	\label{fig:6}
\end{figure}

\subsubsection{Circuit of zero filling for high frequency coefficient}

Fig. 7 shows the circuit of zero filling for high frequency coefficient. The size of the original image is $H\times W$, and $h+w$ qubits are used to store the position information, $q$ qubits are used to store the color information. To get the middle image $\left| \psi  \right\rangle $, we generate a new image size of $H\times W$. The color information of the upper left corner is the low frequency coefficient and the remaining is ${{\left| 0 \right\rangle }^{\otimes q}}$, it is encoded by $\left| C_{0}^{'} \right\rangle \left| C_{1}^{'} \right\rangle ...\left| C_{q-1}^{'} \right\rangle $. The position information of the generated image $\left| \psi  \right\rangle $ is encoded by $\left| X_{0}^{'} \right\rangle \left| X_{1}^{'} \right\rangle ...\left| X_{w-1}^{'} \right\rangle $ and $\left| Y_{0}^{'} \right\rangle \left|Y_{1}^{'} \right\rangle ...\left| Y_{h-1}^{'} \right\rangle $. The generation process of $\left| \psi  \right\rangle $ can be divided into two steps.

Step 1: Initialize image

In Fig. 7, the initial state ${{\left| 0 \right\rangle }^{\otimes h+w+q}}$ is prepared firstly. Then, $h+w$ Hadamard gates are used to change  ${{\left| 0 \right\rangle }^{\otimes h+w}}$ to $\frac{1}{{{\sqrt{2}}^{h+w}}}\sum\limits_{{Y}'=0}^{{{2}^{h}}-1}{\sum\limits_{{X}'=0}^{{{2}^{w}}-1}{{{\left| 0 \right\rangle }^{\otimes q}}\left| {Y}'{X}' \right\rangle }}$, the initial value of $\left| C_{0}^{'} \right\rangle \left| C_{1}^{'} \right\rangle ...\left| C_{q-1}^{'} \right\rangle $ keeps ${{\left| 0 \right\rangle }^{\otimes q}}$ unchanged. In addition, $\left| {{X}_{0}} \right\rangle \left| {{X}_{1}} \right\rangle ...\left| {{X}_{{w}'-1}} \right\rangle $ and $\left| {{Y}_{0}} \right\rangle \left| {{Y}_{1}} \right\rangle ...\left| {{Y}_{{h}'-1}} \right\rangle $ represent the position information of the low frequency image $\left| {{I}_{1}} \right\rangle $, where, ${h}'=h-1$ and ${w}'=w-1$.
${{\left| {{C}_{0}} \right\rangle }^{\prime }}...{{\left| {{C}_{q-1}} \right\rangle }^{\prime }}$ represent the color information of $\left| {{I}_{1}} \right\rangle $. 

Step 2: Set the color information to the low frequency coefficient

In order to set the color information of the low position of $\left| \psi  \right\rangle $ ($\left| X_{0}^{'} \right\rangle =0$ and $\left| Y_{0}^{'} \right\rangle =0$) to the corresponding low frequency coefficient, we use ETOF gates to select every pixel of $\left| {{I}_{1}} \right\rangle $, and assign it to the corresponding pixel of $\left| \psi  \right\rangle $. Each block in the figure represents one pixel is assighed, such block needs ${{2}^{h-1}}\times {{2}^{w-1}}$ in total.

After the two steps, the intermediate image is obtained, it stores the low frequency coefficient at the low position ($\left| X_{0}^{'} \right\rangle =0$ and $\left| Y_{0}^{'} \right\rangle =0$), and the remaining positions have a color value of ${{\left| 0 \right\rangle }^{\otimes q}}$.

\begin{figure}[h]
	\centering
	\includegraphics[width=0.8\textwidth]{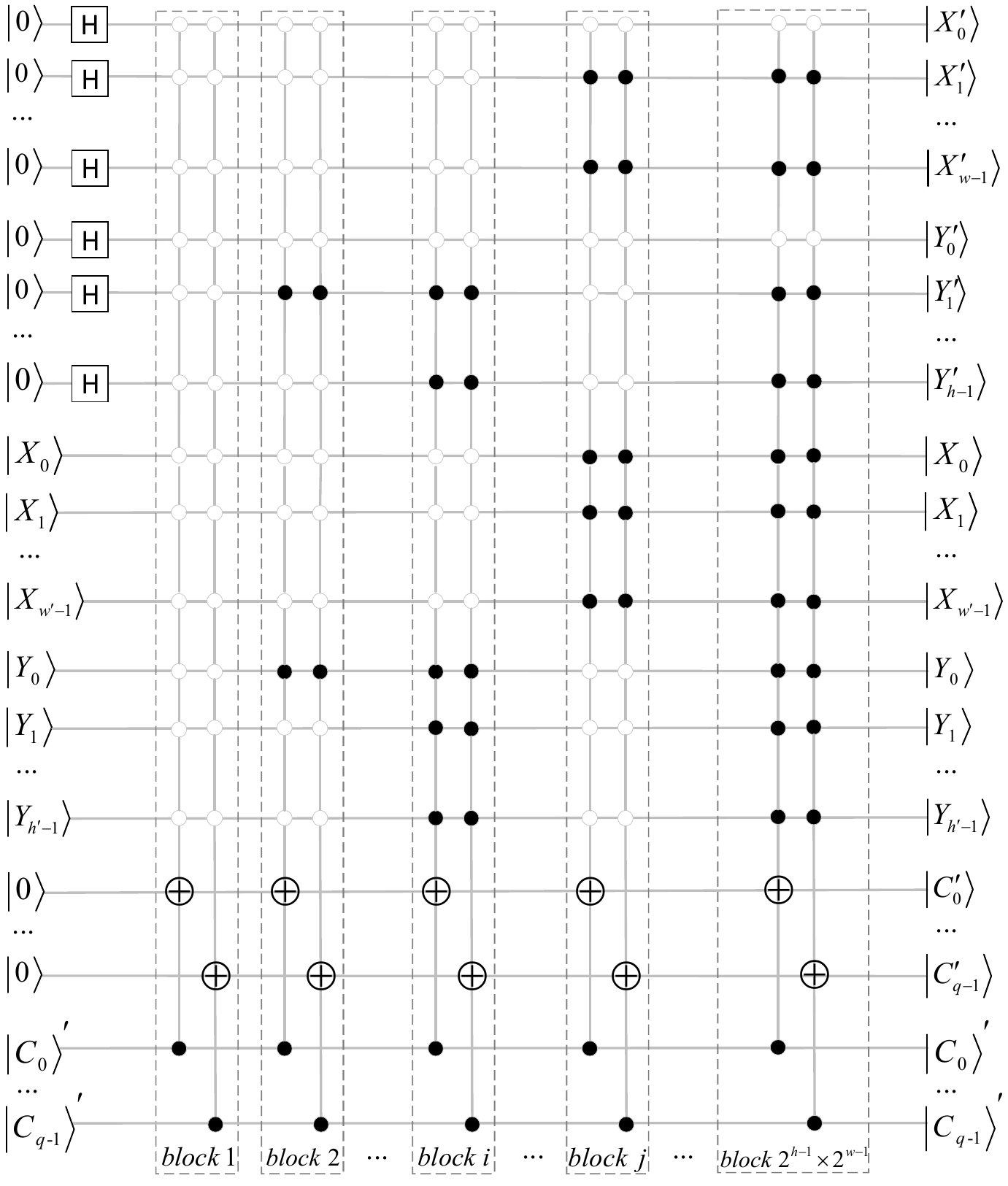}
	\caption{Circuit of zero filling for the high frequency coefficient }
	\label{fig:7}
\end{figure}

\subsubsection{Circuits of the encryption and decryption processes}

Fig. 8 gives the complete circuit of the encryption process. QWT is the quantum wavelet transform module, OLC is the obtaining low frequency information module, and XOR is the XOR operation module, where, the circuits of QWT and XOR have been given in Ref. \cite{17,7}. After QWT, the plain image $\left| I \right\rangle $ becomes a wavelet frequency domain image $\left| T \right\rangle $, and then the low frequency part $\left| {{I}_{1}} \right\rangle $ is output through the OLC module, finally the XOR module is used to obtain the encrypted image $\left| Q \right\rangle $.

\begin{figure}[h]
	\centering
	\includegraphics[width=0.8\textwidth]{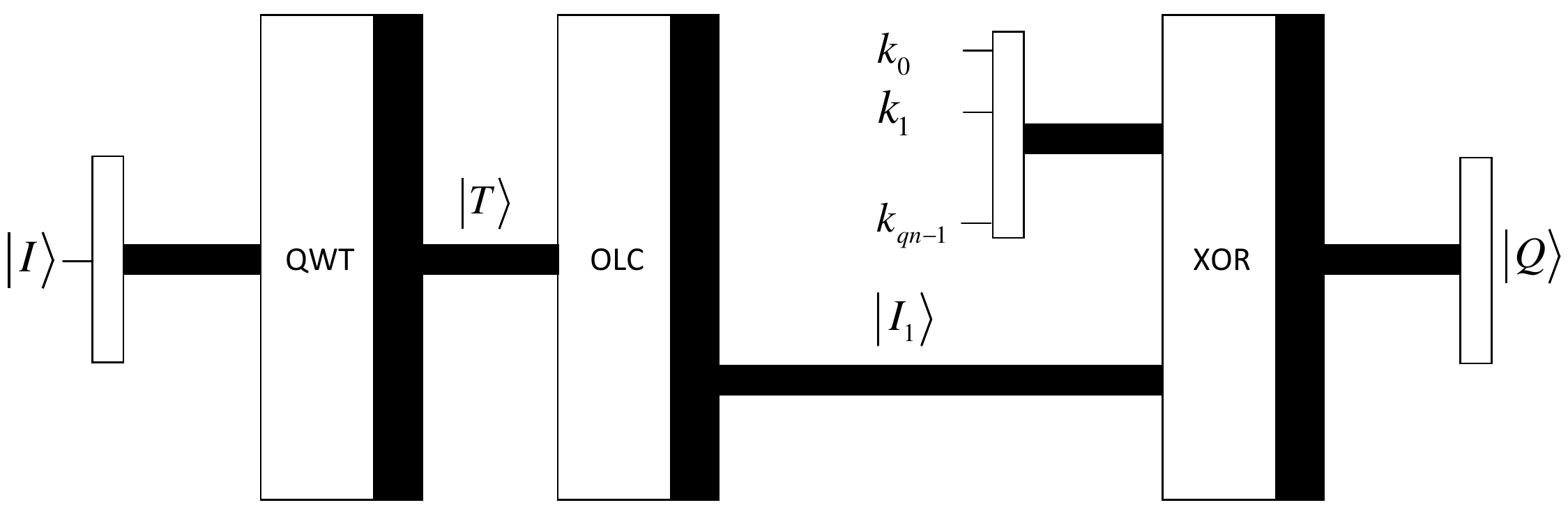}
	\caption{Circuit of the encryption process}
	\label{fig:8}
\end{figure}

Fig. 9 shows the complete circuit of the decryption process. ZFH is the zero filling for high frequency coefficient module, IQWT is the inverse quantum wavelet transform module. After XOR module, the cipher image $\left| Q \right\rangle$ becomes a low frequency image $\left| {{I}_{1}} \right\rangle$, then $h+w+q$ initial states $\left| 0 \right\rangle $ and image $\left| {{I}_{1}} \right\rangle$ are as input of ZFH module to obtain the intermediate image $\left| \psi  \right\rangle $. Finally, the decrypted image $\left| {{I}'} \right\rangle $ is obtained through the IQWT module.

\begin{figure}[h]
	\centering
	\includegraphics[width=0.8\textwidth]{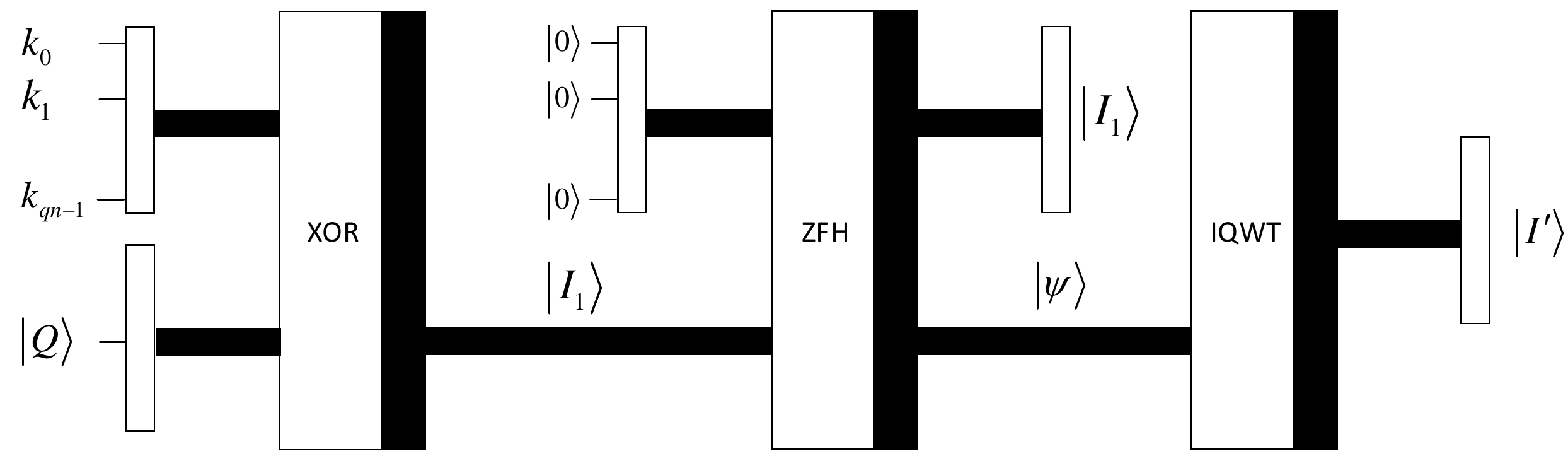}
	\caption{Circuit of the decryption process}
	\label{fig:9}
\end{figure}

\section{Experiment and Analysis}  

In this section, it would give a detailed numerical and theoretical analyses of the encryption algorithm in this paper. The numerical simulations use the MATLAB in a classical computer. It selects the plain images “boat”, “peppers”, and “plane” with size of $\text{512}\times \text{512}$. It uses logistic map to generate the key stream, carries out the image decomposition by Daubechies’ D4 wavelet transform. The encrypted and decrypted images are shown in Fig. 10, we can see that the encrypted images can hide the plaintext information effectively, and the decrypted images can recover the original information accurately.

\begin{figure}[!h]
	\centering
	\begin{minipage}{0.32\textwidth}
		\centering
		\includegraphics[width=\textwidth]{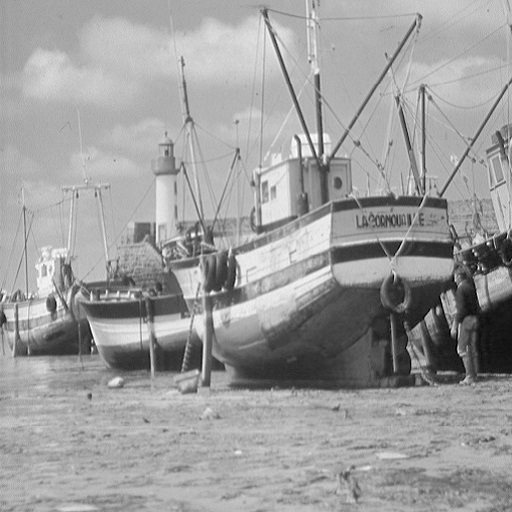}\\
		(1)
	\end{minipage}
	\begin{minipage}{0.32\textwidth}
		\centering
		\includegraphics[width=\textwidth]{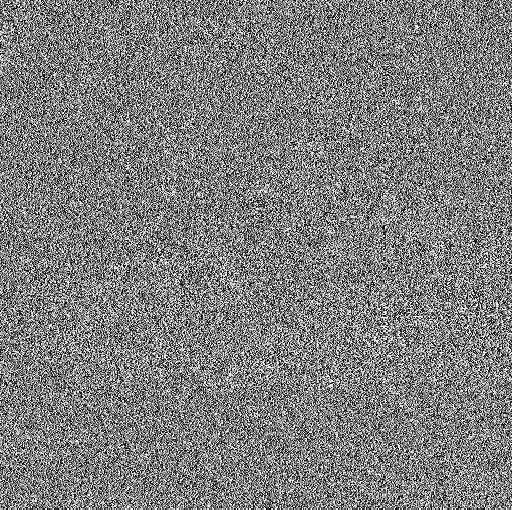}\\
		(2)
	\end{minipage}
	\begin{minipage}{0.32\textwidth}
		\centering
		\includegraphics[width=\textwidth]{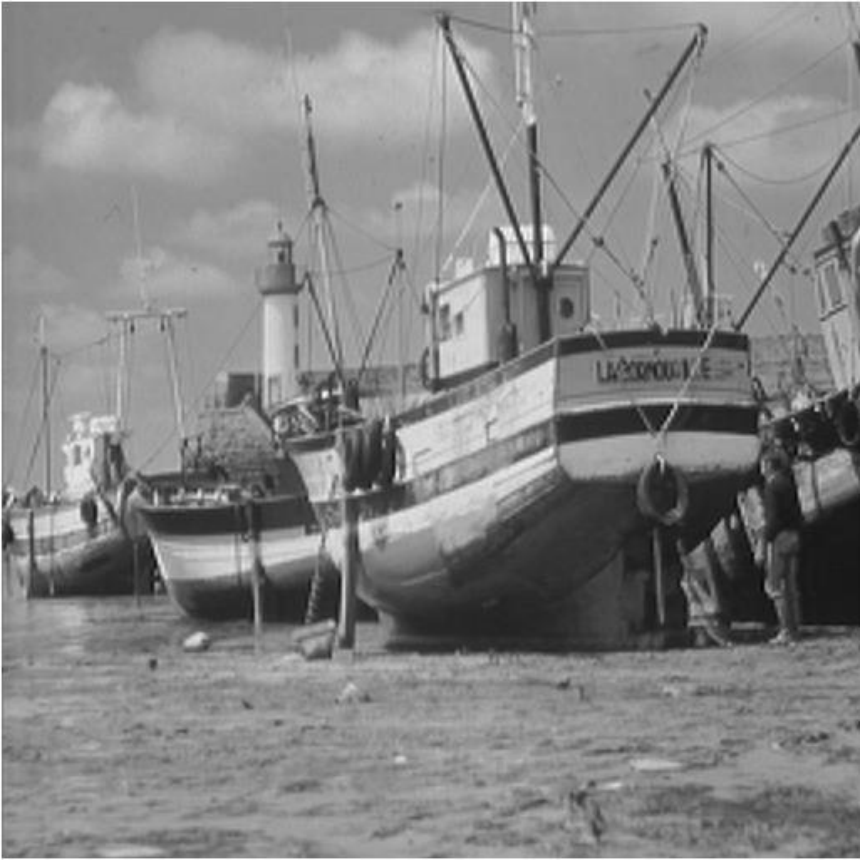}\\
		(3)
	\end{minipage}
\begin{minipage}{0.32\textwidth}
	\centering
	\includegraphics[width=\textwidth]{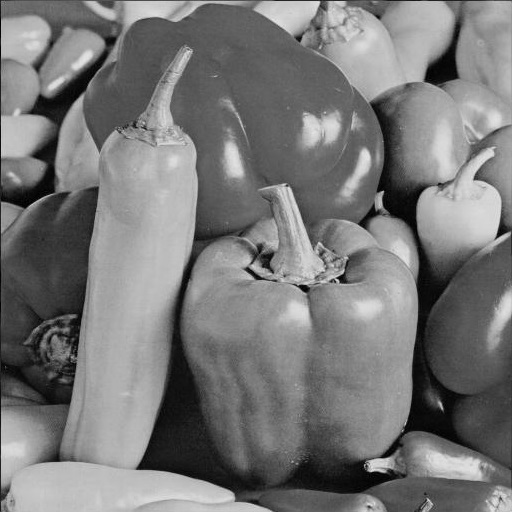}\\
	(4)
\end{minipage}
\begin{minipage}{0.32\textwidth}
	\centering
	\includegraphics[width=\textwidth]{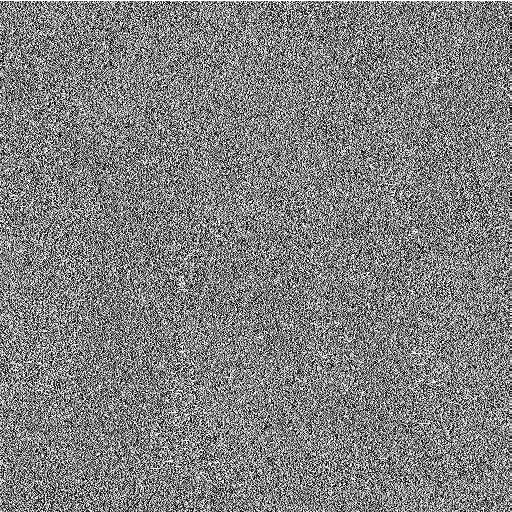}\\
	(5)
\end{minipage}
\begin{minipage}{0.32\textwidth}
	\centering
	\includegraphics[width=\textwidth]{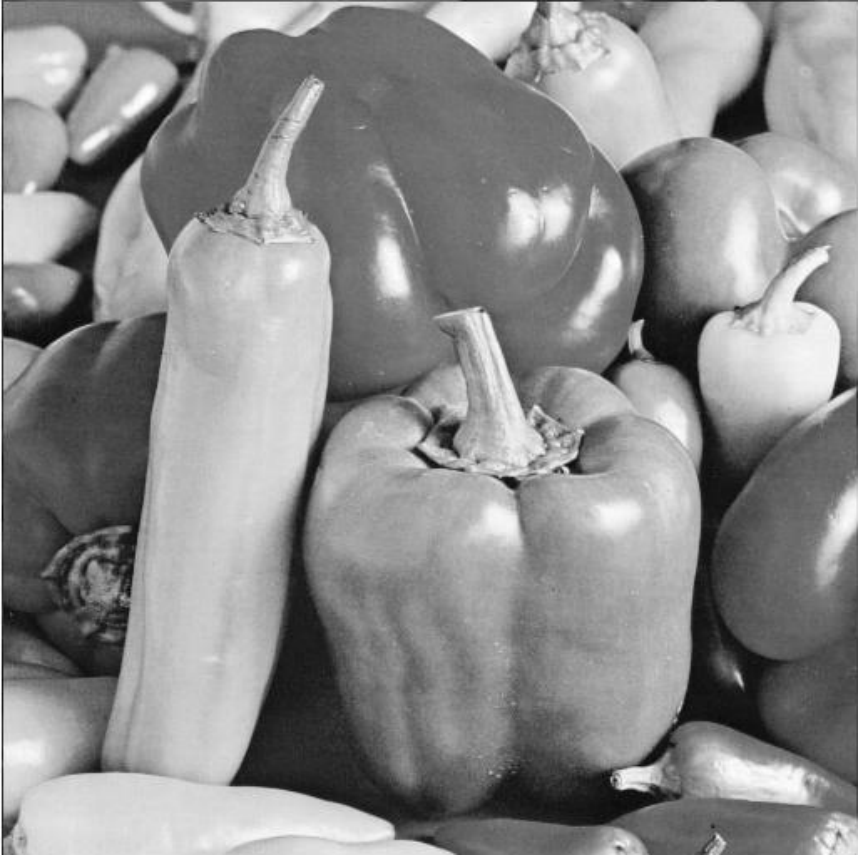}\\
	(6)
\end{minipage}
\begin{minipage}{0.32\textwidth}
	\centering
	\includegraphics[width=\textwidth]{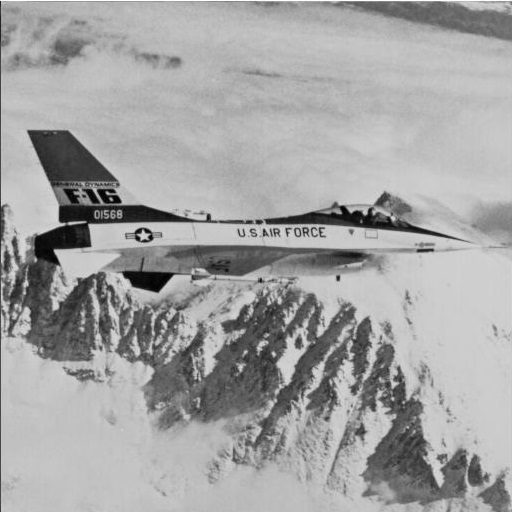}\\
	(7)
\end{minipage}
\begin{minipage}{0.32\textwidth}
	\centering
	\includegraphics[width=\textwidth]{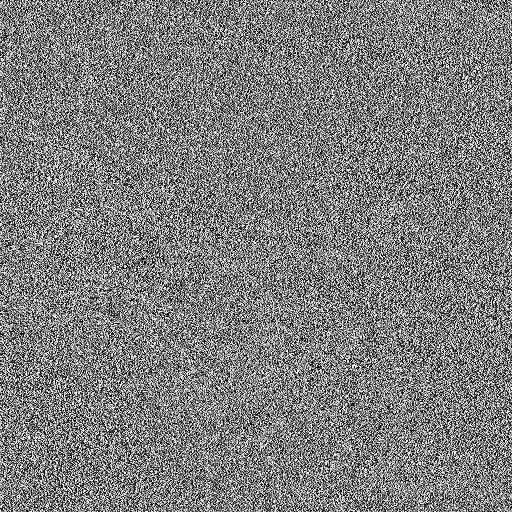}\\
	(8)
\end{minipage}
\begin{minipage}{0.32\textwidth}
	\centering
	\includegraphics[width=\textwidth]{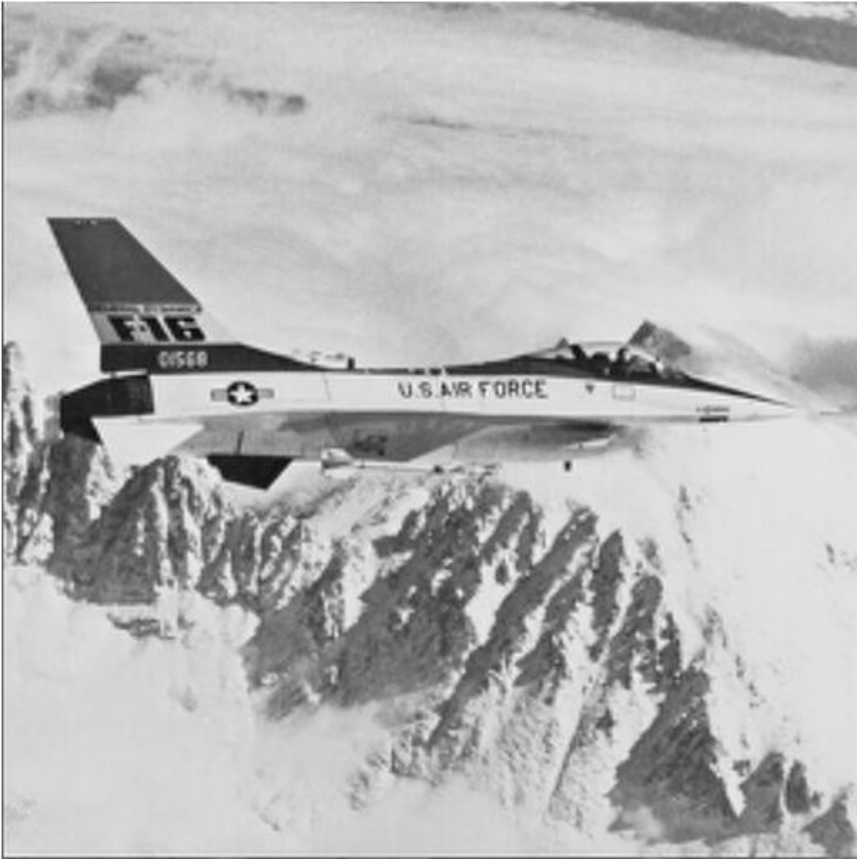}\\
	(9)
\end{minipage}
	\caption{ (1),(4),(7): Plain images of “boat”, “peppers” and “plane”; (2),(5),(8): Encrypted images of “boat”, “peppers” and “plane”;   (3),(6),(9): Decrypted images of “boat”, “peppers” and “plane”
	}
\end{figure}

\subsection{Statistical Analysis}

It is very necessary to make the statistical analysis for the cipher images. In order to verify if the quantum image encryption algorithm based on wavelet transform can reach to the ideal encryption and decryption effect, we will analyze from five aspects: PSNR value of image, correlation of adjacent pixels, information entropy, spatial frequency and histogram.

\subsubsection{PSNR of Image}

PSNR (Peak Signal to Noise Ratio)  is the objective criterion for measuring the degree of image distortion. The larger PSNR value between the plain image and the decrypted image, the more similar the two images are.The general criterion of PSNR is 30db, computing formula is shown as below.
\begin{equation}
PSNR=\text{10}\log \left( \frac{MA{{X}^{2}}}{MSE} \right)
\end{equation}
\begin{equation}
MSE=\frac{\text{1}}{HW}\sum\limits_{i=1}^{H}{\sum\limits_{j=1}^{W}{{{\left\| K\left( i,j \right)-I\left( i,j \right) \right\|}^{2}}}}
\end{equation}
$MAX$ represents the maximum value of image color, $MSE$ represents the mean square error, namely the mean square error between image $I$ and $K$ size of $H\times W$(original image and processed image) . From Table 1, we can see that PSNR value of this algorithm is very closely to or larger than the criterion 30db, therefore, the degree of distortion of the final decrypted image is within the acceptable scope, it has a good visual effect.

\renewcommand\arraystretch{1.5}
\begin{table}[h]
	\centering
	\caption{The PSNR value of decrypted images}
	\begin{tabular}{m{40pt}<{\centering}m{60pt}<{\centering}}
		\hline
		Image & PSNR \\
		\hline
		Boat & 29.4898\\
		Peppers & 32.4578\\
		Plane & 31.8846\\
		\hline
	\end{tabular}
\end{table}
\subsubsection{Correlation of Adjacent Pixels} 
The correlation of adjacent pixels is an important statistical indicator of image \cite{20}. In order to observe visually the change of correlation of the image before and after the encryption, the distribution of the horizontal, vertical and diagonal direction of plain images and cipher images are shown in Fig. 11. The adjacent pixels of plain image is usually close to each other. Hence, in Fig. 11(1), (3) and (5), most of the points are centered around a 45-degree slash. However, an effective encryption algorithm should ensure that the correlation between the two adjacent pixels of the encrypted image is close to zero. Therefore, all points in Fig. 11(2), (4) and (6) are evenly distributed in the rectangular region.

In order to discuss and analyze the adjacent pixels of the encryption algorithm, the correlation of the adjacent pixels in the horizontal, vertical and diagonal directions of the images are calculated respectively. The definition of correlation coefficients is shown below.
\begin{equation}
{{r}_{xy}}=\frac{E\left( \left( x-E\left( x \right) \right)\left( y-E\left( y \right) \right) \right)}{\sqrt{D\left( x \right)D\left( y \right)}}
\end{equation}
In the formula, $x$ and $y$ are gray values of two adjacent pixels in a image, $E\left( x \right)$ and $D\left(x\right)$ respectively represent the expectation and variance of $x$. From Table 2, we can see that the correlation values of the adjacent pixels on all directions of plain images are close to 1, it shows that the adjacent pixels of plain image has a very strong correlation. However, the correlation values of the adjacent pixels of the corresponding cipher images are close to the ideal value 0, it shows the correlation of the pixels of cipher images are weaker. Therefore, this encryption algorithm can weaken the correlation of the adjacent pixels of plain images effectively. Hence, an attacker could not exploit the dependency.

\begin{figure}[!h]
	\centering
	\begin{minipage}{0.48\textwidth}
		\centering
		\includegraphics[width=\textwidth]{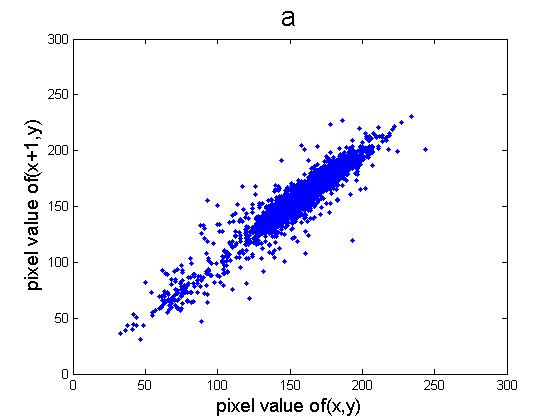}\\
		(1)
	\end{minipage}
	\begin{minipage}{0.48\textwidth}
		\centering
		\includegraphics[width=\textwidth]{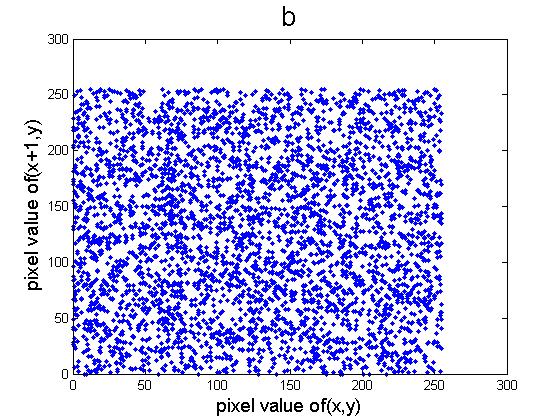}\\
		(2)
	\end{minipage}
	\begin{minipage}{0.48\textwidth}
		\centering
		\includegraphics[width=\textwidth]{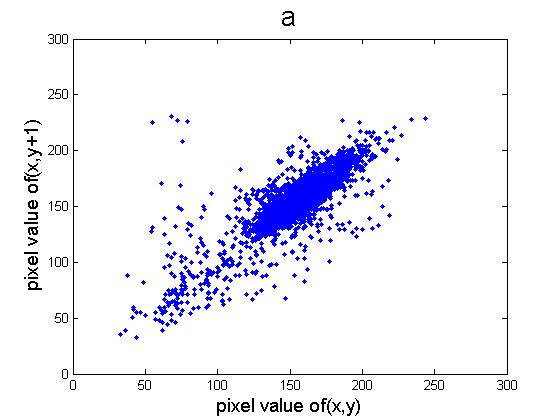}\\
		(3)
	\end{minipage}
	\begin{minipage}{0.48\textwidth}
		\centering
		\includegraphics[width=\textwidth]{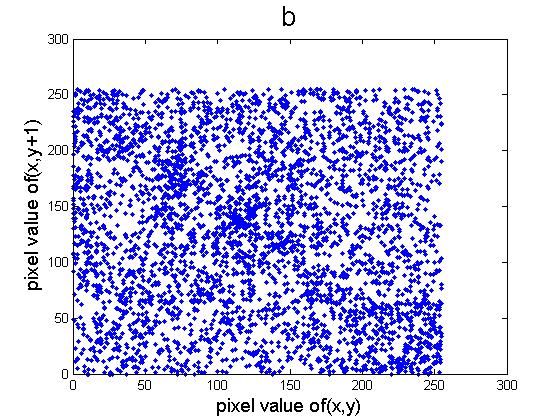}\\
		(4)
	\end{minipage}
	\begin{minipage}{0.48\textwidth}
		\centering
		\includegraphics[width=\textwidth]{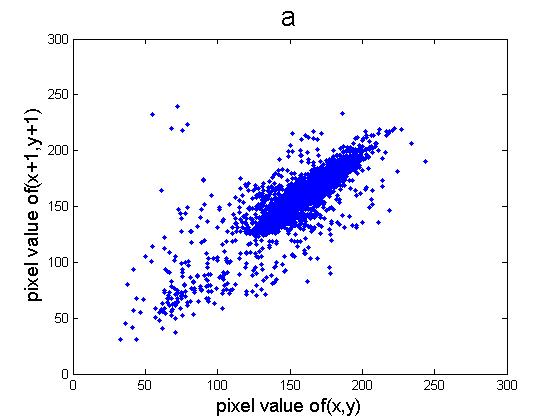}\\
		(5)
	\end{minipage}
	\begin{minipage}{0.48\textwidth}
		\centering
		\includegraphics[width=\textwidth]{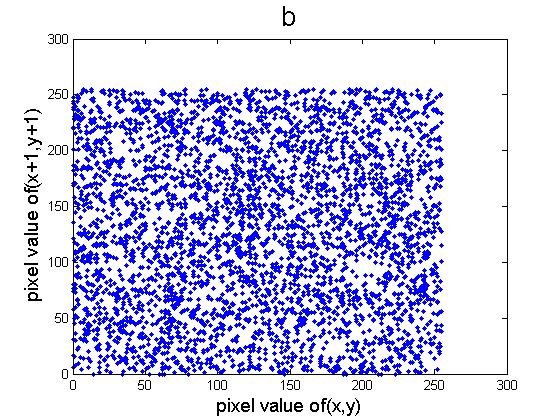}\\
		(6)
	\end{minipage}
	\caption{ Correlation distributions of two adjacent pixels: (1)horizontal adjacent pixels of “boat”; (2)horizontal adjacent pixels of encrypted “boat”; (3)vertical adjacent pixels of “boat”; (4)vertical adjacent pixels of encrypted “boat”; (5)diagonal adjacent pixels of “boat”; (6)diagonal adjacent pixels of encrypted “boat”
	}
\end{figure}

\begin{table}[h]
	\centering
	\caption{Correlation of the image}
	\begin{tabular}{cm{35pt}m{35pt}m{35pt}m{38pt}m{38pt}m{38pt}}
		\hline\noalign{\smallskip}
		Images & Horizontal direction of the original image & Vertical direction of the original image & Diagonal
		direction of the original image & Horizontal direction of the encrypted image & Vertical direction of the encrypted image & Diagonal
		direction of the encrypted image\\
		\noalign{\smallskip}\hline\noalign{\smallskip}
		Boat & 0.944259 & 0.827848 & 0.828004 & -0.013823 & -0.106776 & 0.013414\\
		Peppers & 0.976446 & 0.965057 & 0.945684 & -0.007309 & -0.086897 & -0.006581\\
		Plane & 0.962170 & 0.969954 & 0.947365 & 0.001593 & -0.103145 & -0.003359\\
		\noalign{\smallskip}\hline
	\end{tabular}
\end{table}

\subsubsection{Information Entropy} 
Information entropy is used to represent the uncertainty of image information, if the gray values of a image are distributed more even, and then the entropy value is more larger. The computing formula of information entropy is shown as below,
\begin{equation}
H\left( S \right)=-\sum\limits_{i=0}^{{{2}^{n}}-1}{p\left( {{s}_{i}} \right){{\log }_{2}}p\left( {{s}_{i}} \right)}
\end{equation}
where, $S$ represents the gray-scale assemblage, $p\left( {{s}_{i}} \right)$ represents the probability of the occurrence of the $ith$ gray level.

Table 3 gives out the information entropy of plain and cipher images. By the analysis ,we find out the entropy value of the encrypted plain images are close to the ideal value 8, it can be approximately thought that the cipher images will not cause the information leakage. Therefore, this encryption algorithm can resist the information entropy attacks, and has a very good security.

\begin{table}[h]
	\centering
	\caption{The information entropy}
	\begin{tabular}{c>{\centering}m{100pt}>{\centering\arraybackslash}m{100pt}}
		\hline\noalign{\smallskip}
		Images & The information entropy of the original image & The information entropy of the cipher image\\
		\noalign{\smallskip}\hline\noalign{\smallskip}
		boat & 7.191370 & 7.993751\\
		peppers & 7.592451 & 7.996350\\
		plane & 6.705888 & 7.992652\\
		\noalign{\smallskip}\hline
	\end{tabular}
\end{table}
\subsubsection{Spatial Frequency}  
The spatial frequency of an image reflects the overall activity of the spatial domain. For an image with $M$ rows $N$ columns, the formula of its spatial frequency is show as below.
\begin{equation}
SF=\sqrt{{{\left( RF \right)}^{2}}+{{\left( CF \right)}^{2}}}
\end{equation}
$RF$ represents the row frequency of image, $CF$ represents the column frequency of image. Formula (21) and (22) respectively give out the definition of $RF$ and $CF$,where $F\left( x,y \right)$ is the gray value on position $\left( x,y \right)$. Therefore, we can see that $RF$ and $CF$ respectively reflect the changes of image on the horizontal direction and vertical direction.

Table 4 gives out the numerical values of $RF$, $CF$ and $SF$ of images, we can see that three indexes of the encrypted images are greater than the indexes of the plain image. Hence, the randomness of the encrypted image is perfect.
\begin{equation}
RF=\sqrt{\frac{1}{M\times N}\sum\limits_{x=0}^{M-1}{\sum\limits_{y=0}^{N-1}{{{\left[ F\left( x,y \right)-F\left( x,y-1 \right) \right]}^{2}}}}}
\end{equation}
\begin{equation}
CF=\sqrt{\frac{1}{M\times N}\sum\limits_{x=0}^{M-1}{\sum\limits_{y=0}^{N-1}{{{\left[ F\left( x,y \right)-F\left( x\text{-1},y \right) \right]}^{2}}}}}
\end{equation}

\begin{table}[!h]
	\centering
	\caption{Spatial frequency}
	\begin{tabular}{c>{\centering}m{30pt}>{\centering}m{30pt}>{\centering}m{30pt}>{\centering}m{40pt}>{\centering}m{40pt}>{\centering\arraybackslash}m{40pt}}
		\hline\noalign{\smallskip}
		Images & RF of original image & CF of original image & SF of original image & RF of encrypted image & CF of encrypted image & SF of encrypted image\\
		\hline\noalign{\smallskip}
		Boat & 16.405837 & 11.178274 & 19.8521 & 108.511421 & 102.535811 & 149.2927\\
		Peppers & 11.595972 & 10.980018 & 15.9696 & 109.198045 & 103.501510 & 150.4552\\
		Plane & 12.023420 & 12.417315 & 17.2845 & 110.598764 & 105.796779 & 153.0524\\
		\noalign{\smallskip}\hline
	\end{tabular}
\end{table}

\subsubsection{Histogram}
Histogram is another important statistical characteristic of image, it is usually used to analyze the image encryption algorithm performance. Different plain images have the different gray-scale histograms. 

The histograms of plain images in this paper are shown as Fig. 12(1),(3) and (5), and they are quite different. Fig. 12(2), (4) and (6) show the corresponding gray scale histograms of the encrypted images, and they are very similar and close to the ideal distribution. Therefore, this algorithm can resist the attackers to obtain the useful information from the histograms of the encrypted images.

\begin{figure}[!h]
	\centering
	\begin{minipage}{0.48\textwidth}
		\centering
		\includegraphics[width=\textwidth]{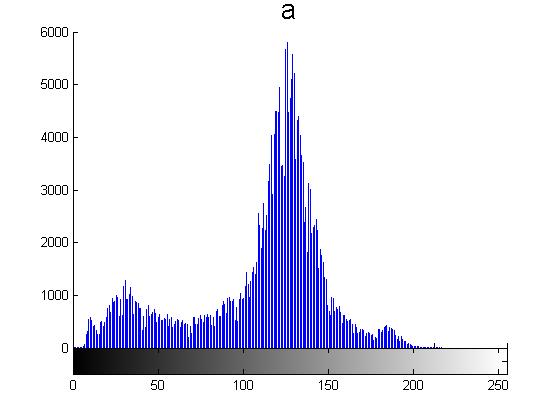}\\
		(1)
	\end{minipage}
	\begin{minipage}{0.48\textwidth}
		\centering
		\includegraphics[width=\textwidth]{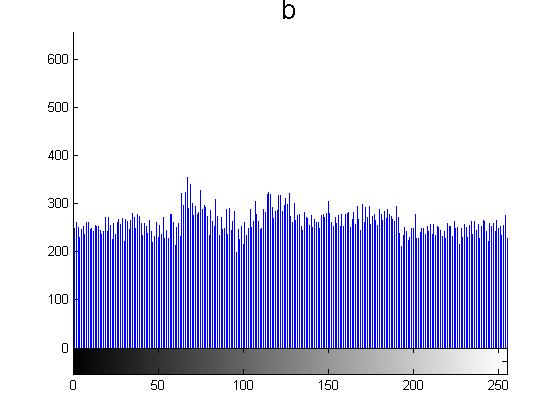}\\
		(2)
	\end{minipage}
	\begin{minipage}{0.48\textwidth}
		\centering
		\includegraphics[width=\textwidth]{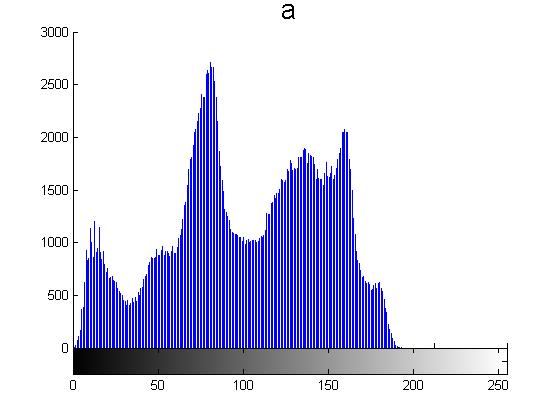}\\
		(3)
	\end{minipage}
	\begin{minipage}{0.48\textwidth}
		\centering
		\includegraphics[width=\textwidth]{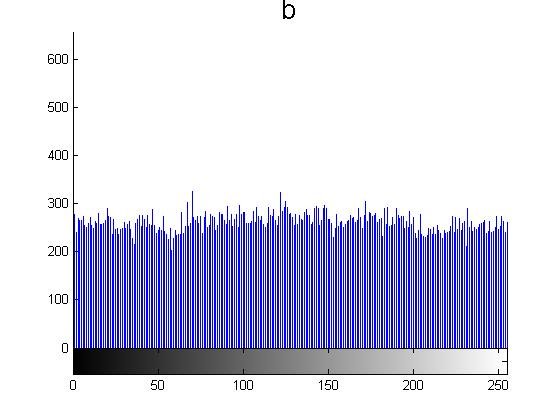}\\
		(4)
	\end{minipage}
	\begin{minipage}{0.48\textwidth}
		\centering
		\includegraphics[width=\textwidth]{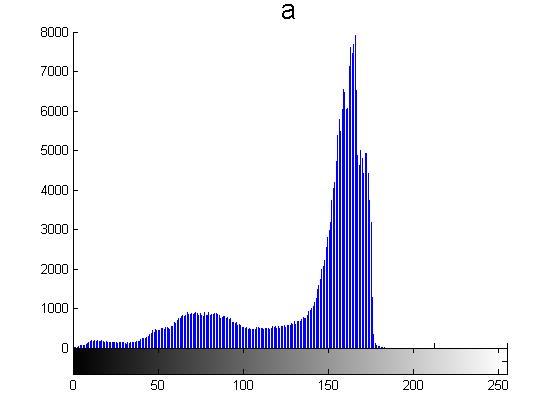}\\
		(5)
	\end{minipage}
	\begin{minipage}{0.48\textwidth}
		\centering
		\includegraphics[width=\textwidth]{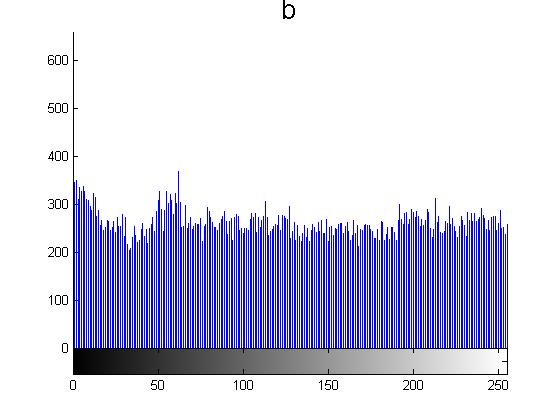}\\
		(6)
	\end{minipage}
	\caption{Histograms: (1) “boat”, (2) encrypted “boat”, (3) “peppers”, (4) encrypted “peppers”, (5) “plane”, (6) encrypted “plane”
	}
\end{figure}

\subsection{Security Analysis}
The image encryption algorithm with better encryption effect should have enough key space to resist the brute-force attack. In the proposed algorithm, if the attacker doesn’t know the parameters of the logistic map which used to generate the key stream, then the key space of this algorithm will rely on the generated key stream. As we know, the length of the key stream is equal to the size of the plain image. Suppose the size of the plain image is $512\times 512$ and every position uses 8 bits to represent the gray value, then the total length is $512\times 512\times 8$. Thus, the key space of this algorithm is ${{2}^{512\times 512\times 8}}$, which is ${{2}^{{{2}^{21}}}}$ and large enough to resist the brute-force attack.

\subsection{Computational Complexity Analysis}
The complexity of the algorithm mentioned by this paper is mainly divided into two parts: One is the complexity of quantum wavelet transform, another is the complexity of XOR operation. References \cite{16,17} have carried out the detailed analysis for the complexity of quantum wavelet transform, from it, we can get the complexity of the first-class Daubechies’ D4 quantum wavelet transform is $O\left( n \right)$. At the same time, from Ref. \cite{21}, we know for a ${{\text{2}}^{n}}\times {{2}^{n}}$ image, its XOR operation needs $\text{384}n-384$ basic quantum gates, and this paper just encrypts the low frequency information (1/4 original image), so the time complexity of XOR operation in this algorithm is linear. In conclusion, the total computing complexity of the quantum image encryption algorithm based on wavelet transform shall be $O\left( n \right)$. 

To illustrate the efficiency of the proposed algorithm, we compare the complexity of the proposed algorithm with the existing algorithms. Firstly, for references \cite{11,22}, their complexity is much higher than that of the proposed algorithm, reaching the exponential level, because they perform transformations for too many times. Secondly, for references \cite{4,23}, their complexity depends on the complexity of quantum Fourier transform, so the complexity of those algorithms is at least $O\left( {{n}^{2}} \right)$, larger than the proposed algorithm based on wavelet transform. Although Ref. [24] also achieves encryption by performing wavelet transform, this algorithm only regards wavelet transform as a common frequency domain transformation, and does not make full use of the characteristics of wavelet transform. Therefore, compared with the algorithm in this paper, the complexity of this algorithm in Ref. \cite{24} is still very large. In conclusion, our method has a certain improvement on the encryption complexity.

\section{Conclusion}

This paper puts forward an adaptive quantum image encryption and decryption algorithm based on wavelet transform which is adaptive to the scenes that need to encrypt a large number of images during network transmission. We reserve the image low frequency information only, so as to reduce the encryption workload, then the encryption process is realized by implementing XOR operation. Statistical simulation demonstrates that the quantum image encryption algorithm has higher security for the encrypted images and lower degree of distortion for the decrypted images. At the same time, the theoretical analysis proves this algorithm has the lower computing complexity than the most existing algorithms. 

It should be pointed out that the algorithm is free from the disadvantage that the encryption depends too much on the frequency of transformations, but it is affected by the performance of the wavelet transform itself. In the next step, we will further improve the efficiency of quantum image encryption by using quantum nature.

\end{document}